\documentclass[11pt]{article}
\usepackage{epsfig}
\usepackage{float}
\usepackage{graphicx}
\usepackage{cite}
\usepackage{appendix}
\usepackage[utf8]{inputenc}
\usepackage[left=25mm,top=25mm,bottom=25mm,right=25mm]{geometry}
\usepackage{multirow}
\usepackage{hhline}
\usepackage{amssymb}
\usepackage{pifont}
\usepackage[footnotesize]{caption} 
\newcommand{\xmark}{\ding{53}}%
\begin{document}
\bigskip

\newcommand{\be}{\begin{equation}}
\newcommand{\en}{\end{equation}}
\newcommand{\bea}{\begin{eqnarray}}
\newcommand{\ena}{\end{eqnarray}}
\newcommand{\noi}{\noindent}
\newcommand{\ra}{\rightarrow}
\newcommand{\bib}{\bibitem}
\newcommand{\refb}[1]{(\ref{#1})}
\newcommand{\bff}{\begin{figure}}
\newcommand{\eff}{\end{figure}}

\begin{center}
{\Large \bf Charged black holes from a family of Born-Infeld-type electrodynamics models}
\end{center}
\hspace{0.0 cm}
\begin{center}
Sebasti\'an Belmar-Herrera\footnote{sebastian.belmar@ufrontera.cl} and Leonardo Balart\footnote{leonardo.balart@ufrontera.cl} \\{\small \it Departamento de Ciencias F\'{\i}sicas, \\
Facultad de Ingenier\'{\i}a y Ciencias \\ \small Universidad de La Frontera, Casilla 54-D \\
Temuco, Chile.}
\end{center}

\hspace{0.3 cm}

\begin{abstract}
We present a family of nonlinear electrodynamics models that are free of the infinite self-energy of the point charge.
Each model is dependent on a dimensional nonlinearity parameter and is determined by the integer value of a  dimensionless parameter $n$. The Born-Infeld model is recovered when $n=1$. Some of the characteristics of this family are studied. 
In addition, we study the solutions of electrically charged AdS black holes, which result from coupling these electrodynamic models to General Relativity. A Smarr formula consistent with the first law of thermodynamics is also obtained in an extended phase space.

\end{abstract}

\hspace{0.8cm}

\noindent{\it Key words}: Static black hole; Born-Infeld type electrodynamics; Smarr formula.


\section{Introduction}
\label{intro}

  
 The model of Born-Infeld electrodynamics~\cite{born} is free of singularities. The electric field derived from this model decreases from its maximum value at the origin (equivalent to the nonlinearity parameter $b$ of the model) until it behaves like the electric field of model Maxwell at large distances.
One can also see that the self-energy of a point charge is finite. The parameter $b$ has been related to the string tension~\cite{Gibbons:2001gy,Fradkin:1985qd}. 
Possible constraints for the parameter $b$ of this model have been analyzed in Refs.~\cite{Davila:2013wba,Ellis:2017edi,NiauAkmansoy:2017kbw,NiauAkmansoy:2018ilv,Neves:2021tbt,Neves:2021jdy,DeFabritiis:2021qib}.

Unlike the Euler-Heisenberg electrodynamics~\cite{Heisenberg:1936nmg}, the Born-Infeld model does not exhibit the phenomenon of vacuum birefringence in the presence of external electric field~\cite{Kruglov:2007bh, Kim:2021grj}. 
The causality and unitarity principles (tachyons and ghosts are absent, respectively) are held in the Born-Infeld theory~\cite{Shabad:2011hf}.

Born-Infeld electrodynamics has inspired other models free of singularities, which, in general, have the same characteristics: the Soleng model~\cite{Soleng:1995kn}, various models presented by S. Kruglov~\cite{Kruglov:2014hpa,Kruglov:2014iwa,Kruglov:2015fcd,Kruglov:2016cdm,Kruglov:2017fuj,Kruglov:2017xmb}, others proposed by Gaete et al.~\cite{Gaete:2013dta,Gaete:2017cpc,Gaete:2018nwq}, the exponential model~\cite{Hendi:2012zz}, the models presented by Mazharimousavi and Halilsoy~\cite{Mazharimousavi:2019sgz,Mazharimousavi:2021uki} and the double-logarithmic model~\cite{Gullu:2020ant}.

Some models that generalize the BI model have been recently introduced, that is, at some limit of the model parameters, the Born-Infeld electrodynamics is recovered: the Born-Infeld like model with two parameter\cite{Gaete:2014nda}, the Born-Infeld type model with three parameters~\cite{Kruglov:2016uzf}, the ModMax model~\cite{Bandos:2020jsw}, the ModMax generalized model~\cite{Kruglov:2021bhs}.


The first extension of the Reissner-Nordstr\"om solution was obtained using Born-Infeld electrodynamics~\cite{Hoffmann:1935ty,Hoffmann:1937noa}. 
In contrast to Bardeen-type black holes (for example Refs.~\cite{AyonBeato:1998ub,AyonBeato:1999rg,Bronnikov:2000vy,Dymnikova:2004zc,Balart:2014jia,Balart:2014cga}) whose metric functions and curvature invariants are regular in everywhere, a charged black hole obtained with electrodynamics of Born-Infeld, presents singularities when evaluating these quantities.
Additionally, depending on the values of the mass, the charge and the parameter $b$ of the Born-Infeld black hole, the solutions obtained can be categorized as Reissner-Nordstr\"om type or Schwarzschild type~\cite{Fernando:2003tz}. The first ones can have two horizons or one degenerate horizon; the second ones have a single horizon.

The energy-momentum tensor of the Born-Infeld model has a non-zero trace, because of this a Smarr formula consistent with the first law of black hole thermodynamics cannot be found without considering a new thermodynamic quantity~\cite{Rasheed:1997ns}  (see also Ref.~\cite{Balart:2017dzt}). 
In the study of the thermodynamics of Born-Infeld Anti-de Sitter black holes in the extended phase space~\cite{Gunasekaran:2012dq}, where the cosmological constant represents the thermodynamic pressure~\cite{Kastor:2009wy, Dolan:2010ha, Dolan:2011xt}, the parameter $b$ has been considered as a thermodynamic variable, which, together with its corresponding conjugate quantity~\cite{Yi-Huan:2010jnv}, allows obtaining a Smarr formula consistent with the first law of thermodynamics. In this framework, the quantity conjugated to the nonlinearity parameter has been interpreted as vacuum polarization per unit volume~\cite{Gunasekaran:2012dq}. Other works~\cite{Zou:2013owa, Hendi:2014kha, Kumar:2019zbp, Zeng:2019jta, Gullu:2020qni, Balart:2021glm, Kumar:2022fyq}  have also considered thermodynamics in the extended phase space, also profiting the nonlinearity parameter of the model as an extra thermodynamic quantity, to study some features of Born-Infeld or Born-Infeld-type black holes.

In this paper we propose a family of nonlinear electrodynamics models that are also dependent on a nonlinearity parameter $b$, but where the different solutions of this family are obtained by specifying the value of a dimensionless parameter
 $n = 1, 2, ...$, which determines the order of the field strength that follow the term obtained from Maxwell electrodynamics when $r \rightarrow \infty$. In this family when $n = 1$ the Born-Infeld model is obtained and in all cases when $b \rightarrow  \infty$ we recover the Maxwell model. We study some of the characteristics that these models present, including the phenomenon of birefringence in the presence of an external magnetic field and later in the presence of an external electric field.  In addition, we obtain static electrically charged AdS black hole solutions by coupling Einstein equations to these electrodynamics models and from here we obtain algebraic expressions for both the electric field and the metric function. 
We also show that for each value of $n$, a black hole solution is obtained that has a different behavior depending on the mass, the electric charge and the parameter $b$, which is manifested in the number of horizons it has, such as is the case for the Born-Infeld black hole. Finally, considering $b$ as a new thermodynamic quantity in the extended phase space allows us to obtain a Smarr formula consistent with the corresponding first law of thermodynamics for any value of $n$.


The paper is organized as follows. In section 2, the family of nonlinear electrodynamics models considered is presented and some of its main features are studied. In section 3, we present the black hole solutions that are obtained by coupling General Relativity with these electrodynamic models. Section 4 is devoted to describe the different types of behavior of black hole solutions for each value of $n$. Section 5 presents the obtaining of a Smarr formula consistent with the first law of black hole thermodynamics. Conclusions are given in section~6.


\section{Family of Born-Infeld-type electrodynamics models}
In this section we propose a family of new nonlinear electrodynamics models. The motivation for introducing this family is that it has two features of interest. First, each of the models in this family is free of singularities in self-energy of the point charge. Second, the experimental interest in measuring the effect of birefringence (see for example Ref.~\cite{Beard:2021dka}), which is present in these models, except for the case $n = 1$.

Let us summarize the family of new nonlinear electrodynamics models in the following Lagrangian
\begin{equation} 
\mathcal{L}(\mathcal{F},\mathcal{G})=-\left(\mathcal{F}-\frac{1}{2 b^2} \mathcal{G}^2 \right) {}_2 F_1 \left(\frac{1}{2n},\frac{1}{n}; 1+\frac{1}{n}; \frac{(-2)^{n}}{b^{2n}} \left(\mathcal{F}-\frac{1}{2 b^2} \mathcal{G}^2 \right)^{n} \right)
\label{L-BI-gral} \, ,
\end{equation}
where $n > 0$ is an integer, $\mathcal{F} = F_{ \mu \nu} F^{ \mu \nu}/4 = (B^2 - E^2)/2$, $\mathcal{G} = F_{ \mu \nu} \tilde{F}^{ \mu \nu}/4 = \mathbf{E} \cdot \mathbf{B}$, $b$ is a nonlinearity parameter of the theory and $_{2}F_1\left(a, b; c; z\right)$ is the Gauss hypergeometric function. In the limit $b\rightarrow \infty$ (in the weak field limit) the Lagrangian behaves as $\mathcal{L} = - \mathcal{F}$.
Note that when $n = 1$ in the Lagrangian~(\ref{L-BI-gral}), we recover the Lagrangian of Born-Infeld electrodynamics~\cite{born}\begin{equation} 
\mathcal{L}_{BI} = b^2 \left(1-\sqrt{1 +\frac{2 \mathcal{F}}{b^2} -\frac{\mathcal{G}^2}{b^4}}\right)
\label{L-BI} \, ,
\end{equation}
The form of Lagrangian~(\ref{L-BI-gral}) is inspired by that of the Born-Infeld model when written using the Gauss hypergeometric function.

If we assume $\mathbf{B} = 0$ and $\mathbf{E} = E(r)\hat{r}$, that is, $\mathcal{F} = -E^2(r)/2$ and $\mathcal{G} = 0$, then from Eq.~(\ref{L-BI-gral}) we obtain the following field equations
\begin{equation} 
\nabla_\mu (F^{\mu\nu} \mathcal{L}_\mathcal{F}) = 0 
\label{electrom-eq-gral}  \, ,
\end{equation} 
where $\mathcal{L}_\mathcal{F}$ denotes differentiation of $\mathcal{L}$ with respect to $\mathcal{F}$,
which yields
\begin{equation}
\frac{E} {\left[1-(-1)^{-n} b^{-2 n}
   \left(-E^2\right)^n\right]^{\frac{1}{2 n}}} = \frac{q}{r^2}
\,\,\label{Eq-E-BI-gral} \,  .
\end{equation}
The solution of this equation provides an expression for the electric field
\begin{equation}
E(r) = \frac{b \, q}{(q^{2n} + b^{2n} r^{4n})^\frac{1}{2n}}
\,\,\label{E-gral} \,  .
\end{equation}

Here, it can be noted that the electric field asymptotically behaves as the Maxwell field, i.e.,
\begin{equation}
E(r) = \frac{q}{r^2} - \frac{q^{2n+1}}{2n\,  b^{2n}\, r^{2(2n+1)}} + \frac{(2 n+1)  q^{4 n+1} }{8 n^2b^{4 n}r^{2 (4 n+1)} } + O\left(\frac{1}{r^{2 (6 n+1)}}\right)
\,\,\label{E-gral-large} \,  .
\end{equation}

\begin{figure}[h]
\centering
\includegraphics[width=0.6\textwidth]{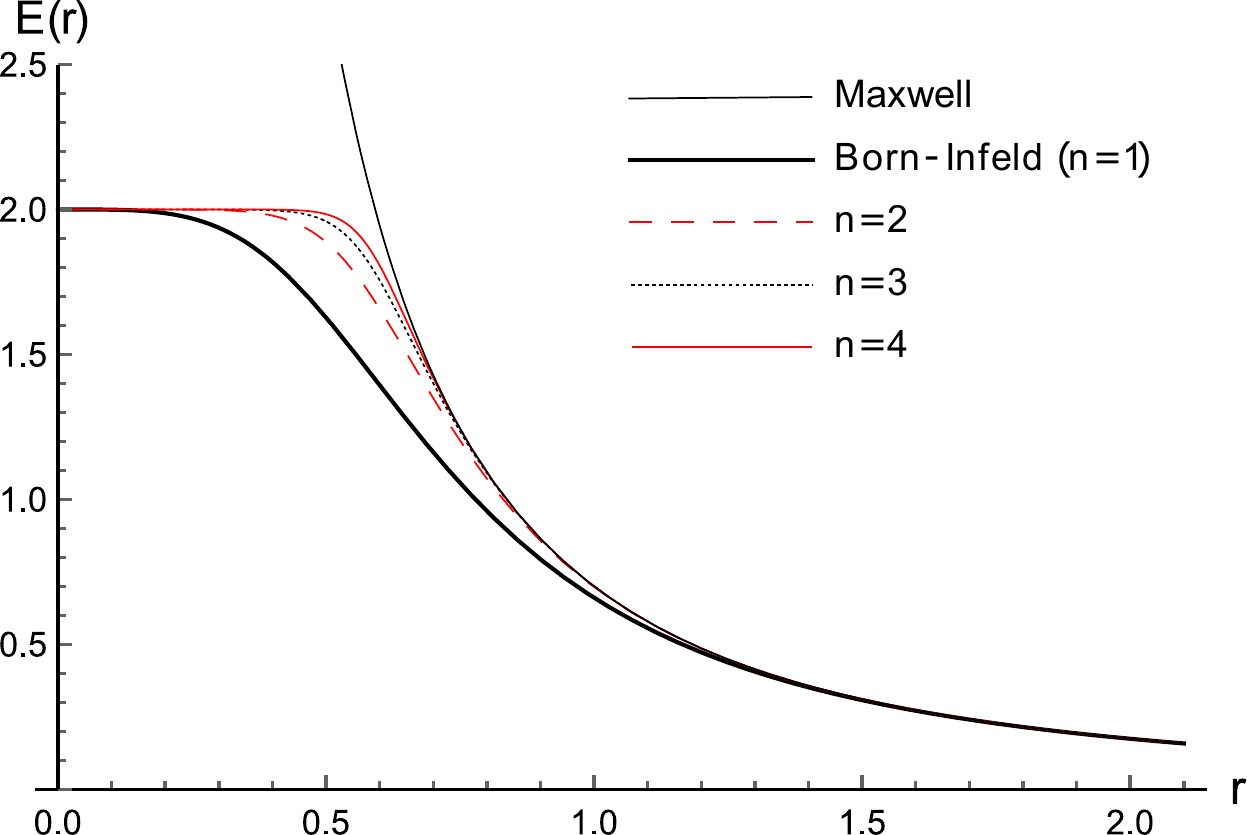}
\caption{The electric field is plotted for some values of $n$. Maxwell electric field is also depicted. Here, in all cases, the parameters are chosen as $q = 0.7$ and $b = 2$.}
\label{fig:Ebig} 
\end{figure}

For its part, expanding the electric field about $r = 0$ yields
\begin{equation}
E(r) = b -\frac{b^{2 n+1} } {2 n q^{2 n}}r^{4 n} + \frac{(2n+1) b^{4 n+1}  }{8 n^2 q^{4 n}}r^{8 n} + O\left(r^{12 n}\right)
\,\,\label{E-gral-small} \,  .
\end{equation}
In addition one can see that the parameter $b$ represents the maximum value of the electromagnetic field strength for any value of $n$, just as is known to occur for the case $n = 1$.
The shape of the electric field~(\ref{E-gral}) for $n = 1, 2, 3$ and $4$ is shown in Fig.~\ref{fig:Ebig} where the starting point is at the value of $b$. For comparison purposes, the figure also shows the Maxwell case.

\subsection{Self-energy}

In Ref.~\cite{Dehghani:2021fwb} the authors conjecture that for theories belonging to the Born-Infeld family in 3+1-dimensions, the value of the self-energy for charged particles is finite. In our case, we calculate these values and obtain a result that is finite for each $n$, as it is shown in the following.

Self-energy in 3+1 dimensions is defined as
\begin{equation}
\varepsilon = 4 \pi \int^{\infty}_{0} T_0^0 \, r^2 dr =
 - \int^{\infty}_{0} \left(E^2 \mathcal{L}_\mathcal{F}+\mathcal{L}\right)\, r^2 dr 
\label{self-energy}     \, .
\end{equation}
Looking at Eq.~(\ref{L-BI-gral}), one readily obtains
\begin{equation}
\mathcal{L}_\mathcal{F}= - \left( 1- \frac{(-2)^{n} \mathcal{F}^n}{b^{2n}}\right)^{-\frac{1}{2 n}}
\label{Der-L}
\end{equation}
which together with the expression for the electric field allows us to calculate the self-energy
\begin{equation}
\varepsilon (n) = \frac{2 \Gamma \left( \frac{1}{4 n}\right) \Gamma \left( \frac{4 n+1}{4 n}\right)}{3 \Gamma \left( \frac{1}{2 n}\right)}\sqrt{b q^3}
\label{self-energy-n}
\end{equation}
where $\varepsilon (1) = 1.23605 \sqrt{b q^3}$ and it can reach a maximum value of $\varepsilon (n\rightarrow \infty ) =  4\sqrt{b q^3}/3$.


\subsection{Causality and unitarity} \label{cau-unit}

Here we analyze the causality and unitarity of the model when $n$ is odd and when it is even.

The principles of causality and unitarity are satisfied if 
$\mathcal{L}_\mathcal{F} \leq 0$, $\mathcal{L}_{\mathcal{FF}} \geq 0$, $\mathcal{L}_{\mathcal{GG}} \geq 0$, $\mathcal{L}_\mathcal{F} + 2 \mathcal{F} \mathcal{L}_{\mathcal{FF}} \leq 0$ and $2 \mathcal{F}\mathcal{L}_\mathcal{GG} - \mathcal{L}_{\mathcal{F}} \geq 0$~\cite{Shabad:2011hf}.

If we consider Lagrangian~(\ref{L-BI-gral})  and the abbreviation  $\mathcal{K}=- 2b^{2} \mathcal{F}+\mathcal{G}^2$, we get 
\begin{equation}
\mathcal{L}_\mathcal{F} = - \frac{1}{ \left (  1-b^{-4 n}\mathcal{K}^{n}\right )^{\frac{1}{2n}}} 
\end{equation}

\begin{equation}
\mathcal{L}_{\mathcal{F}\mathcal{F}}  = \frac{b^4 \mathcal{K}^{n - 1}}{ \left (  b^{4 n}- \mathcal{K}^{n}\right )^{1+\frac{1}{2n}}}
\end{equation}

\begin{equation}
\mathcal{L}_{\mathcal{G}\mathcal{G}} =  \frac{b^{4n} +2 b^2 \mathcal{F}\mathcal{K}^{n-1}}{ \left (  b^{4 n}- \mathcal{K}^{n}\right )^{1+\frac{1}{2n}}} 
\end{equation}

\begin{equation}
\mathcal{L}_\mathcal{F} + 2 \mathcal{F} \mathcal{L}_{\mathcal{F}\mathcal{F}}  = -\frac{b^2\left (  b^{4n} - \mathcal{G}^2 \mathcal{K}^{n-1} \right )}{ \left (  b^{4 n}-\mathcal{K}^{n}\right )^{1+\frac{1}{2n}}} 
\end{equation}

\begin{equation}
  2 \mathcal{F} \mathcal{L}_{\mathcal{G}\mathcal{G}}-\mathcal{L}_\mathcal{F}   = \frac{\left [b^{4n}\left ( b^2+2 \mathcal{F} \right ) -b^2\mathcal{K}^{n}+4 b^2\mathcal{F}^2\mathcal{K}^{n-1} \right ]}{ \left (  b^{4 n}- \mathcal{K}^{n}\right )^{1+\frac{1}{2n}}} 
\end{equation}

According to these results, we obtain the following table showing the cases in which the inequalities are satisfied (\checkmark) or not (\xmark).
\begin{table}[h!] 
\begin{center}
\resizebox{10cm}{!} {\begin{tabular}{|c|c|c|c|c|c|}
\hline 
Condition & $n$ & $B=0$ & $E > B$ & $E = 0$ & $B > E$\\
\hline
\multirow{2}{*}{$\mathcal{L}_\mathcal{F} \leq 0$ } & odd & \checkmark & \checkmark& \checkmark  & \checkmark   \\
\hhline{~-----}                  & even            & \checkmark & \checkmark &\checkmark * & \checkmark *  \\
\hline
\multirow{2}{*}{$\mathcal{L}_{\mathcal{FF}} \geq 0$ } & odd & \checkmark & \checkmark & \checkmark & \checkmark  \\
\hhline{~-----}                  & even            & \checkmark & \checkmark & \xmark & \xmark\\
\hline
\multirow{2}{*}{$\mathcal{L}_{\mathcal{GG}} \geq 0$ } & odd & \checkmark & \checkmark & \checkmark &   \checkmark   \\
\hhline{~-----}                  & even           & \checkmark & \checkmark & \checkmark *  &  \checkmark * \\
\hline
\multirow{2}{*}{$\mathcal{L}_\mathcal{F} + 2 \mathcal{F} \mathcal{L}_{\mathcal{FF}} \leq 0$} & odd & \checkmark & \checkmark & \checkmark & \checkmark  \\
\hhline{~-----}                  & even           & \checkmark & \checkmark &\checkmark * &  \checkmark * \\
\hline
\multirow{2}{*}{$2 \mathcal{F}\mathcal{L}_\mathcal{GG} - \mathcal{L}_{\mathcal{F}} \geq 0$} & odd & \checkmark & \checkmark & \checkmark & \checkmark  \\
\hhline{~-----}                  & even           & \checkmark & \checkmark  & \checkmark *  & \checkmark * \\
\hline\end{tabular}}
\caption{Causality and unitarity. Here $\mathcal{G} = 0$ and  the asterik  indicate that the condition is satisfied only if $\mathcal{F} \leq b^2$.}
\label{Ca-Un}
\end{center}
\end{table}

We can notice in the table above that, just like for Born-Infeld electrodynamics~\cite{Shabad:2011hf}, for all other values of $n$ odd  with $\mathcal{G} = 0$ the principles of causality and unitarity are always satisfied. For its part, for $n$ even, only when $\mathcal{F} < 0$ and considering  $\mathcal{G} = 0$, the conditions are fully satisfied.


\subsection{Birefringence}
Another feature that can be considered for our model is the birefringence~\cite{Battesti:2012hf}. The Born-Infeld model does not present the phenomenon of birefringence~\cite{Boillat:1970gw}.  Some authors have considered the study of the phenomenon of birefringence in different models of non-linear electrodynamics: Euler-Heisenberg theory~\cite{Bialynicka-Birula:1970nlh}, Soleng electrodynamics~\cite{Gaete:2013dta}, Born-Infeld-type electrodynamics~\cite{Kruglov:2016uzf}, Gaete-Helay\"el-Neto model~\cite{Gaete:2017cpc}, inverse electrodynamics~\cite{Gaete:2021ytm}, logarithmic electrodynamics~\cite{Gaete:2018nwq}, Euler-Heisenberg-like electrodynamics~\cite{Gaete:2015qda} and electrodynamics with three parameters~\cite{Kruglov:PLA} and ModMax model~\cite{Bandos:2020jsw, Sorokin:2021tge}.

To analyze this phenomenon in the case of our model, and following Ref.~\cite{Kruglov:2007bh}, we consider the propagation of the electromagnetic plane wave $(\mathbf{E}_p,\mathbf{B}_p)$ in a constant external magnetic field which is along $x$ axis $\overline{\mathbf{B}}=(\overline{B},0,0)$ where
\begin{equation}
\mathbf{E}_p(\textbf{r},t)=\mathbf{E}_{0} \, e^{-i(\omega t - \textbf{k} \cdot \textbf{r})} 
\,\,\, ,  \,\,\,\,\, 
\mathbf{B}_p(\textbf{r},t)=\mathbf{B}_{0} \, e^{-i(\omega t - \textbf{k} \cdot \textbf{r})} \, .
\end{equation}
We assume the wave propagates in the $z$-direction and the background magnetic field is much stronger than the amplitudes of electric and magnetic fields. Then the resultants electromagnetic fields become $\mathbf{B} = \mathbf{B}_p + \overline{\mathbf{B}}$ and $\mathbf{E} = \mathbf{E}_p$. Thus substituting into the Lagrangian~(\ref{L-BI-gral}) we get
\begin{equation}
\mathcal{L}(\mathbf{E}_p,\mathbf{B}_p + \overline{\mathbf{B}})=-\left( F_{\mathbf{B}}\right) \, _2F_1\left(\frac{1}{2 n},\frac{1}{n};1+\frac{1}{n};-2^n \left(F_{\mathbf{B}}\right)^n b^{-2 n}\right)\, ,
\end{equation}
where $F_{\mathbf{B}}$ is defined as
\begin{equation}
F_{\mathbf{B}}=\frac{1}{2} \left ( \frac{\left (\mathbf{B}_p + \overline{\mathbf{B}} \right )^2}{2} -\frac{\mathbf{E}_p^2}{2} \right ) - \frac{\left ( \left ( \mathbf{B}_p + \overline{\mathbf{B}} \right )\cdot \mathbf{E}_p \right )^2}{2 b^2}   \, .
\end{equation}

One can obtain the electric displacement and magnetic field by using $\mathbf{D} = \partial \mathcal{L} / \partial \mathbf{E}$ and  $\mathbf{H} = -\partial \mathcal{L} / \partial \mathbf{B}$,   respectively and taking into account only the linear terms in $\mathbf{E}_p$ and  $\mathbf{B}_p$ 
\begin{equation}
\mathbf{D}_p = \frac{1}{\mathcal{R}} \left(\mathbf{E}_p + \frac{\overline{\mathbf{B}}\, (\overline{\mathbf{B}}\cdot \mathbf{E}_p)}{b^2}\right) 
\label{}
\end{equation}
\begin{equation}
\mathbf{H}_p = \frac{1}{\mathcal{R}} \left(\mathbf{B}_p +\left ( -1 \right )^{n} \frac{\overline{\mathbf{B}}^{2(n-1)} \overline{\mathbf{B}} \, ( \overline{\mathbf{B}}\cdot \mathbf{B}_p)}{ b^{2 n} \mathcal{R}^{2 n}}\right)         \, ,
\label{}
\end{equation}
where we have introduced the abbreviation
\begin{equation}
\mathcal{R} = \left(1 +\left ( -1 \right )^{n} \frac{\overline{\mathbf{B}}^{2 n}}{b^{2 n}} \right)^{\frac{1}{2 n}}  \, .
\label{}
\end{equation}

We may calculate the electrical permittivity and the inverse of the magnetic permeability tensors, using the relations $D_i = \varepsilon_{i j} (E_p)_j$ y $H_i=\left(\mu^{-1}\right)_{ij} (B_p)_j$
\begin{equation}
\varepsilon_{i j}=\frac{1}{\mathcal{R}} \left(\delta_{i j} + \frac{\overline{B}_{i} \overline{B}_{j}}{b^2}\right) 
\label{}
\end{equation}
\begin{equation}
\left(\mu^{-1} \right)_{i j}=\frac{1}{\mathcal{R}} \left(\delta_{i j} +\left ( -1 \right )^{n} \frac{\overline{\mathbf{B}}^{2(n-1)} \overline{B}_{i} \overline{B}_{j}}{ (b \, \mathcal{R})^{2 n}}\right)      \, .
\label{}
\end{equation}
From Maxwell equations written in terms $\mathbf{k}$ and $\omega$, we obtain the wave equation for the electric field $\mathbf{E}_p$~\cite{Kruglov:2007bh}
\begin{equation}
\left[\varepsilon_{i j l} \, \varepsilon_{a b c} \, k_j \left( \mu^{-1} \right)_{l a} k_b + \omega^2 \varepsilon_{i c} \right] (E_p)_c = 0   \,   ,
\label{}
\end{equation}
where $\varepsilon_{i j l}$ is the Levi-Civita symbol.

Substituting $\varepsilon_{ij}$ and $\left(\mu^{-1} \right)_{ij}$ in this last equation, we get the following equation in matrix notation,
\begin{equation}
\frac{\omega^2}{\mathcal{R}}\left [ A I +C  \overline{\textbf{B}} \cdot \overline{\textbf{B}}\right ] \textbf{E}_p=0 \,
\label{}
\end{equation}
where $I$ is the identity matrix,
\begin{equation}
A= 1-n^2-(-1)^{n} \frac{n^2\overline{\mathbf{B}}^{2n}}{b^{2n} \mathcal{R}^{2n}}  \,\,\,\, , \,\,\,\,\,\,\,\,
C=  \frac{1}{b^2}+(-1)^{n}\frac{\overline{\mathbf{B}}^{2(n-1)}}{b^{2n} \mathcal{R}^{2n}}
\label{}
\end{equation}
and 
\begin{equation}
\overline{\textbf{B}} \cdot \overline{\textbf{B}}= \left(
\begin{array}{ccc}
 \overline{B}_x^2 & \overline{B}_x \overline{B}_y & \overline{B}_x \overline{B}_z \\
 \overline{B}_x \overline{B}_y & \overline{B}_y^2 & \overline{B}_y \overline{B}_z \\
 \overline{B}_x \overline{B}_z & \overline{B}_y \overline{B}_z & \overline{B}_z^2 \\
\end{array}
\right)
\label{}  \, .
\end{equation} 
 Then we can find the following eigenvalues
\begin{equation}
\lambda_{1}=1-n^{2}\left(1+(-1)^{n}\frac{\overline{\mathbf{B}}^{2 n}}{(b \mathcal{R})^{2 n}}\right), \,\,\,\, \lambda_2=1- n^2+ \frac{\overline{\mathbf{B}}^2}{b^2}  \, .
\label{}
\end{equation}
By setting $\lambda_1=0$ y $\lambda_2=0$, we arrive at the indices of refraction when the polarization of the electromagnetic wave is perpendicular and parallel to the external magnetic field $\overline{\mathbf{B}}$, respectively
\begin{equation}
n_{\perp} = \sqrt{1-\left ( -1 \right )^{n}\frac{\overline{\mathbf{B}}^{2n}}{ b^{2n}}} \,  , \,\,\,\, \,n_{\parallel}= \sqrt{1+\frac{\overline{\mathbf{B}}^{2}}{b^2}}  \, ,
\label{}
\end{equation}
or if we consider $b$ large we can write them as
\begin{equation}
n_{\perp} \approx 1-(-1)^{n}\frac{\overline{\mathbf{B}}^{2n}}{2 b^{2 n}} \,  , \,\,\,\,  n_{\parallel} \approx 1+\frac{\overline{\mathbf{B}}^2}{2 b^2}
\label{n approx}     \, .
\end{equation}
Note that when $n \neq 1 $ the phenomenon of birefringence is present. In other words, electromagnetic waves polarized in perpendicular directions travel at different velocities. Since the velocity of the electromagnetic wave can be found using the relation $v_{\parallel(\perp)}= c/n_{\parallel(\perp)}$ (here we explicitly write the constant $c$), it can also be noted, even when the higher order contributions are very small, that if $n$ is odd then $n_{\perp} > 1$, whereas if $n$ is even we would have non-physics values, that is $n_ {\perp} < 1$. \\


Analogously, we can now calculate the refractive indices in the presence of only a purely external electric field. If we consider that the direction of the background electric field is along the $x$ axis $\overline{\mathbf{E}}=(\overline{E},0,0)$ and the propagation direction of the wave in the $z$ direction as in the previous case, then in this case we arrive at the following refractive indices
\begin{equation}
n_{\perp}= \left(1-\frac{\overline{\mathbf{E}}^2}{b^2} \right)^{-\frac{1}{2}}  \,  , \,\,\,\,   n_{\parallel}=\left(1-\frac{\overline{\mathbf{E}}^{2 n}}{b^{2 n}} \right)^{-\frac{1}{2}}  \, .
\label{}
\end{equation}
If we consider $b$ large, then
\begin{equation}
n_{\perp} \approx 1+\frac{\overline{\mathbf{E}}^2}{2 b^2}   \,  , \,\,\,\,   n_{\parallel} \approx 1+\frac{\overline{\mathbf{E}}^{2n}}{2 b^{2 n}}  \, .
\label{n approx-E}
\end{equation}
We can notice that, unlike the previous case, the refractive indices are greater than 1 for $n$ even or odd. Here also the phenomenon of birefringence is present, except for the case where $n = 1$.


\section{AdS Black holes with Born-Infeld-type electrodynamics}
In this and the next two sections we present solutions of electrically charged AdS black holes that are obtained from the nonlinear electrodynamics model discussed above. We study the three categories of solutions that are obtained for each value of $n$. We also consider thermodynamics in extended phase space and obtain a Smarr formula consistent with the first law of thermodynamics.

The action of the 3+1-dimensional Einstein theory coupled with a nonlinear electrodynamics model is described by the action
\begin{equation} 
S = \int d^4x \sqrt{-g} \left[ \frac{(R - 2 \Lambda)}{16 \pi G} + \mathcal{L(F)} \right] \, .
\label{action}
\end{equation}
Here, $g$ is the determinant of the metric tensor, $R$ is the Ricci scalar, $G$ is the gravitational constant, $\Lambda = -\frac{3}{l^2}$ is the cosmological constant and $\mathcal{L(F)}$ is the Lagrangian given in Eq.~(\ref{L-BI-gral}).

The Einstein fields equations resulting from action~(\ref{action}) are given by
\begin{equation} 
G_{\mu\nu} + \Lambda g_{\mu\nu} = 8 \pi T_{\mu\nu}
\label{einstein-eq-1}
\end{equation}
\begin{equation} 
4 \pi T_{\mu\nu} = g_{\mu\nu}  \mathcal{L(F)} - F_{\mu\rho} F_\nu^{\,\, \rho} \mathcal{L}_\mathcal{F}
\label{einstein-eq-2} \, 
\end{equation}
and
\begin{equation} 
\nabla_\mu (F^{\mu\nu} \mathcal{L}_\mathcal{F}) = 0
\label{electrom-eq} 
\end{equation}

Let us consider static spherically symmetric black holes where the line elements read
\begin{equation}
ds^2 = -f(r)dt^2 + f(r)^{-1} dr^2 + r^2(d\theta^2 + \sin^2 \theta d\phi^2) \, . 
\end{equation}

Using the Einstein equations~(\ref{einstein-eq-1}) and the Lagrangian~(\ref{L-BI-gral}) allows us to obtain the metric function given by
\begin{eqnarray}
f(r) &=& 1 -  \frac{2 M}{r} + \frac{r^2}{l^2} + \frac{4 \sqrt{bq^3} }{3 r}  
\frac{\Gamma \left(\frac{1}{4n}\right) \Gamma \left(\frac{4n+1}{4n} \right)}{\Gamma\left( \frac{1}{2n}\right)} +  
\frac{2b^2 r^2}{3} \frac{\Gamma \left(\frac{1}{n}\right) \Gamma \left(\frac{2n-1}{2n} \right)}{\Gamma\left( \frac{1}{2n}\right) }
 \nonumber
\\&&
 - \frac{2b q}{3}   \, _{2}F_1\left(-\frac{1}{2n},\frac{1}{2n};1 - \frac{1}{2n} ;-\frac{b^{2n} r^{4n}}{q^{2n}}\right) \nonumber
\\&&  - \frac{4 b q}{3}  \, _{2}F_1\left(\frac{1}{4n},\frac{1}{2n};1 + \frac{1}{4n} ;-\frac{b^{2n} r^{4n}}{q^{2n}}\right) 
\,\,\label{metric-f-gral} \, ,
\end{eqnarray}
where $M$ represents the ADM mass, $q$ the electric charge of the black hole.
This function can also be written as
\begin{eqnarray}
f(r) &=& 1 -  \frac{2 M}{r} + \frac{r^2}{l^2} +
\frac{2b^2 r^2}{3} \frac{\Gamma \left(\frac{1}{n}\right) \Gamma \left(\frac{2n-1}{2n} \right)}{\Gamma\left( \frac{1}{2n}\right) }
 \nonumber
\\&&
 - \frac{2b q}{3}   \, _{2}F_1\left(-\frac{1}{2n},\frac{1}{2n};1 - \frac{1}{2n} ;-\frac{b^{2n} r^{4n}}{q^{2n}}\right) \nonumber
\\&&  + \frac{4 q^2}{3 r^2}  \, _{2}F_1\left(\frac{1}{4n},\frac{1}{2n};1 + \frac{1}{4n} ;-\frac{q^{2n}}{b^{2n} r^{4n}}\right) 
\,\,\label{metric-f-gral-b} \,  \, .
\end{eqnarray}
It can be seen that the asymptotic behavior of this metric function is
\begin{equation} 
f(r) = 1 + \frac{r^2}{l^2}  - \frac{2 M}{r} +  \frac{q^2}{r^2} + O\left(\frac{1}{r^{4 n + 2}}\right)
\label{asymp-RN} \, ,
\end{equation}

\begin{figure}[h]
\centering
\includegraphics[width=0.65\textwidth]{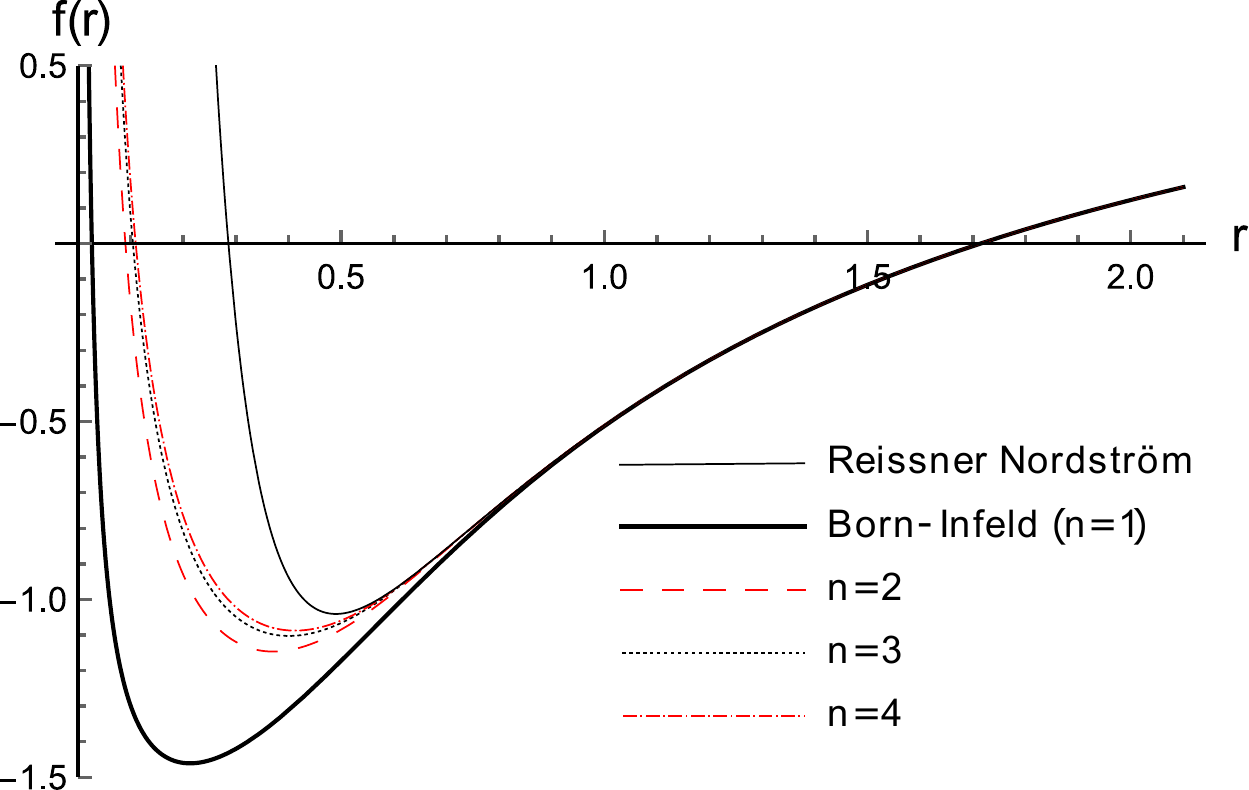}
\caption{Various metric functions corresponding to different values of $n$ are plotted. As in the Reissner-Nordstr\"{o}m and Born-Infeld cases ($n=1$), the solutions for other values of $n$ can also have two horizons, for values of $q$ less than those listed in table~\ref{tab-charge-ex}. However, as in the Born-Infeld case, the solutions can also have only one horizon, depending on the value of $q$, as illustrated below. Here, in all cases, the parameters are chosen as $M = 1$, $q = 0.7$, $\Lambda = 0$ and $b = 2$.}\label{fig:fbig} 
\end{figure}
In Fig.~\ref{fig:fbig}, the metric functions for some values of $n$ together with the Reissner-Nordstr\"{o}m metric function are displayed. 
\begin{figure}[h!]
\centering
\includegraphics[scale=0.5]{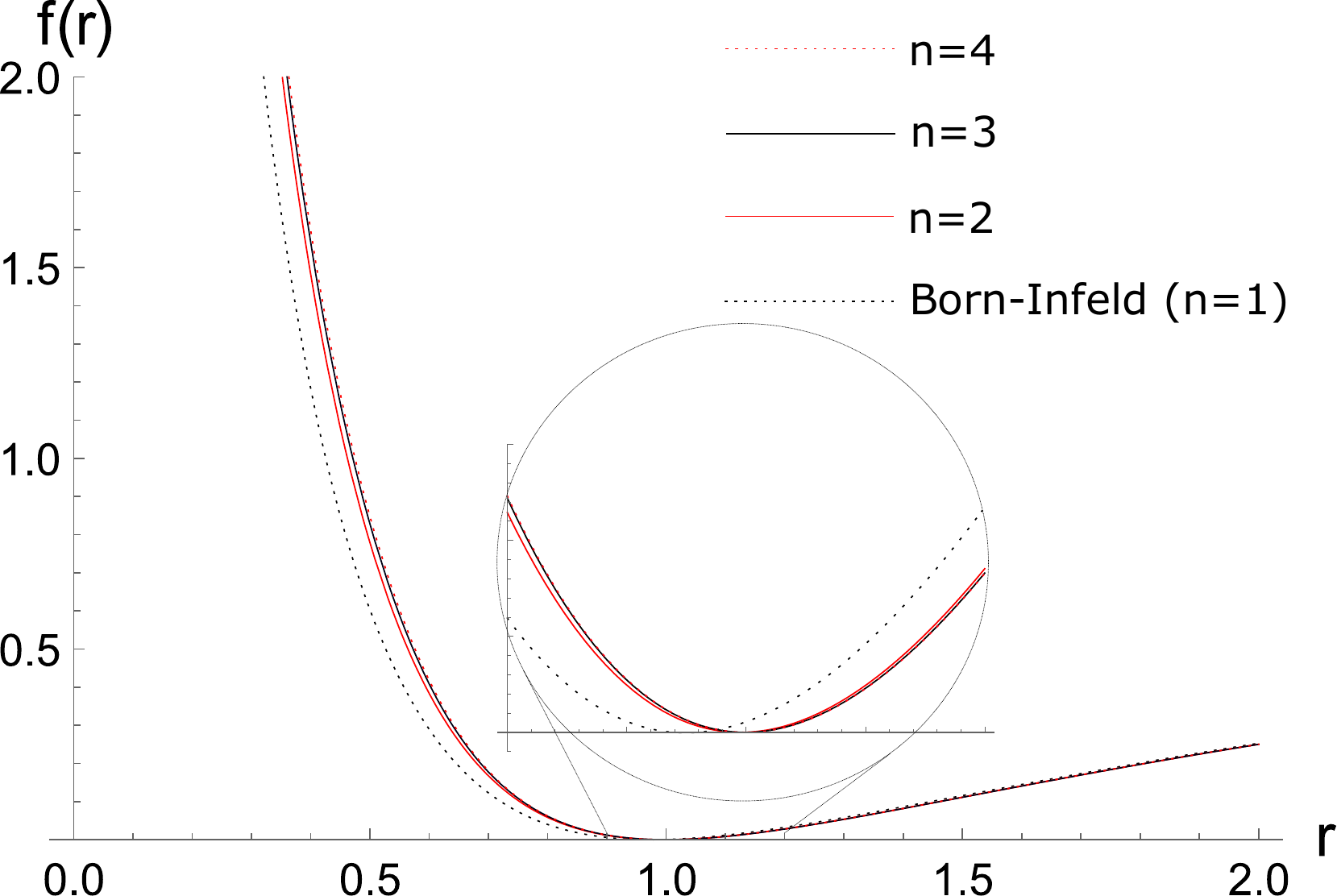}  
\caption{Metric functions of extremal cases for $n=1,2,3$ and $4$. The circle amplifies the region where the outer and inner horizons coincide for the values of $n$ that we are displaying.
In all cases $M = 1$, $\Lambda=0$ and $b=2$, for the electric charge we use the values indicated in table~\ref{tab-charge-ex}.}
\label{BIGcargaextrema}
\end{figure}

It is straightforward to find the mass function $m(r)$ from the metric function, this allows us to obtain
\begin{equation}
\lim_{r \rightarrow 0} m(r) = M - \frac{2 \sqrt{b q^3}  \, \Gamma \left(1+\frac{1}{4 n}\right) \Gamma \left(\frac{1}{4 n}\right)}{3 \Gamma
   \left(\frac{1}{2 n}\right)}    \, .
\label{}
\end{equation}
To ensure that this last result is positive, it is necessary that the electric charge $q$ satisfies the following relation
\begin{equation}
q \leq   \frac{\left(\frac{3}{2}\right)^{2/3} M^{2/3} \Gamma \left(\frac{1}{2 n}\right)^{2/3}}{\sqrt[3]{b} \, \Gamma
   \left(1+\frac{1}{4 n}\right)^{2/3} \Gamma \left(\frac{1}{4 n}\right)^{2/3}} = q_{ex}\, ,
\label{carga_permitida}
\end{equation}
where $q_{ex}$ is the electric charge of the extremal black hole. In Fig.~(\ref{BIGcargaextrema}) extremal solutions are shown for different values of $n$.
\begin{table}[h!]
\centering
\resizebox{4cm}{!}{
\begin{tabular}{|l|l|} 
\hline
\multicolumn{2}{|l|}{Extremal electric charge}  \\ 
\hline
$n$ & \hspace{0.9 cm} $q_{ext}$                          \\ 
\hline
1 & 1.00626550322001                           \\ 
\hline
2 & 1.00028890439369                           \\ 
\hline
3 & 1.00002502213484                           \\ 
\hline
4 & 1.00000287157123                           \\
\hline
\end{tabular}}
\caption{Electric charge of the extremal black holes \\  $q_{ext}$ for $n=1,2,3,4$, with $\Lambda=0$, $M=1$ and $b=2$.}
\label{tab-charge-ex}
\end{table}


\section{Categories of solutions for each value of $n$.}

As we shall see,  for each value of $n$ the metric function can have zero, one or two horizons, depending on the values of $M$, $q$ and $b$. Here we follow the analysis as given in Ref.~\cite{Fernando:2003tz} or Refs.~\cite{Fernando:2006gh, Breton:2017hwe}.

To classify the black hole solutions that can be obtained, it is useful to define the extreme mass. 
For this purpose, let us consider the metric function~(\ref{metric-f-gral}) with fixed values of $M$, $l$ and $n$. Then one can  numerically find the extremal values $q_{ext}$ y $r_{ext}$ for which $f(r) = 0$ and $f^{\prime}(r) = 0$ at the horizon. Hence, the extremal mass of the black hole is defined as
\begin{eqnarray}
M_{ext}(q) &=& \frac{b^2 r^3_{ext} \Gamma \left(1-\frac{1}{2 n}\right) \Gamma \left(\frac{1}{n}\right)}{3 \Gamma \left(\frac{1}{2 n}\right)}+\frac{2 \sqrt{b q^3} \, \Gamma \left(1+\frac{1}{4 n}\right) \Gamma \left(\frac{1}{4 n}\right)}{3 \Gamma
   \left(\frac{1}{2 n}\right)}
   \nonumber
\\ && +\frac{1}{2} \left(\frac{r^3_{ext}}{l^2} + r_{ext}\right) - \frac{b q r_{ext}}{3}  \bigg[ \, _{2}F_1\left(-\frac{1}{2 n}, \frac{1}{2 n};1-\frac{1}{2 n};-\frac{b^{2n} r^{4n}_{ext}}{q^{2n}}\right)
 \nonumber
\\&& + 2 \,  _{2}F_1\left(\frac{1}{4 n}, \frac{1}{2 n};1+\frac{1}{4 n};-\frac{b^{2n} r^{4n}_{ext}}{q^{2n}}\right)\bigg]
    \,  .
\label{M-ext}
\end{eqnarray}
Note that $M_{ext}(q_{ext}) = M$.

At the turn, in the limit $r \rightarrow 0$, the metric function~(\ref{metric-f-gral}) behaves as
\begin{equation}
f(r) = 1 - \frac{2}{r} \left( M - \frac{2\sqrt{b q^3}}{3} \frac{\Gamma \left(\frac{1}{4 n}\right) \Gamma \left(\frac{4 n+1}{4 n}\right)}{\Gamma \left(\frac{1}{2 n}\right)} \right)
- 2 b q + \frac{r^2}{l^2} 
+ \frac{2b^2 r^2}{3} \frac{\Gamma \left(\frac{1}{n}\right) \Gamma \left(\frac{2 n-1}{2 n}\right)}{\Gamma \left(\frac{1}{2 n}\right)} + O(r^{4 n}) \,  .
\label{expa-fm}
\end{equation} 
and using the definition
\begin{equation}
\alpha \equiv \frac{2\sqrt{b q^3}}{3} h(n) \,  ,
\label{alpha}
\end{equation} 
where 
\begin{equation}
h(n)  \equiv \frac{\Gamma \left(\frac{1}{4 n}\right) \Gamma \left(\frac{4 n+1}{4 n}\right)}{\Gamma \left(\frac{1}{2 n}\right)} \,  ,
\label{h-funtion}
\end{equation}
with $1.854 < h(1) < h(2) < h(3) ....  < 2$, one finds the following possible cases:

\vspace{3mm}

\noindent If $M > \alpha$ the solution behaves as a Schwarzschild black hole (S-type) and has only one horizon.

\vspace{3mm}
\noindent If $M < \alpha$ the solution behaves as a Reissner-Nordstr\"{o}m black hole (RN-type) and this can have:

\noindent - a naked singularity, when also $M < M_{ext}$.

\noindent - two horizon, when also $M > M_{ext}$. 

\noindent - a degenerate horizon when also $M = M_{ext}$.

\vspace{3mm}

\noindent If $M = \alpha$, then $f(0) = 1 - 2 b q$ and one can distinguish two cases: 

\noindent - the solution behaves as a Schwarzschild black hole and has only one horizon, when $bq > 1/2$.

\noindent - the solution has a naked singularity, when $bq < 1/2$.
\vspace{3mm}

\begin{figure}[h!]
\centering
\includegraphics[scale=0.8]{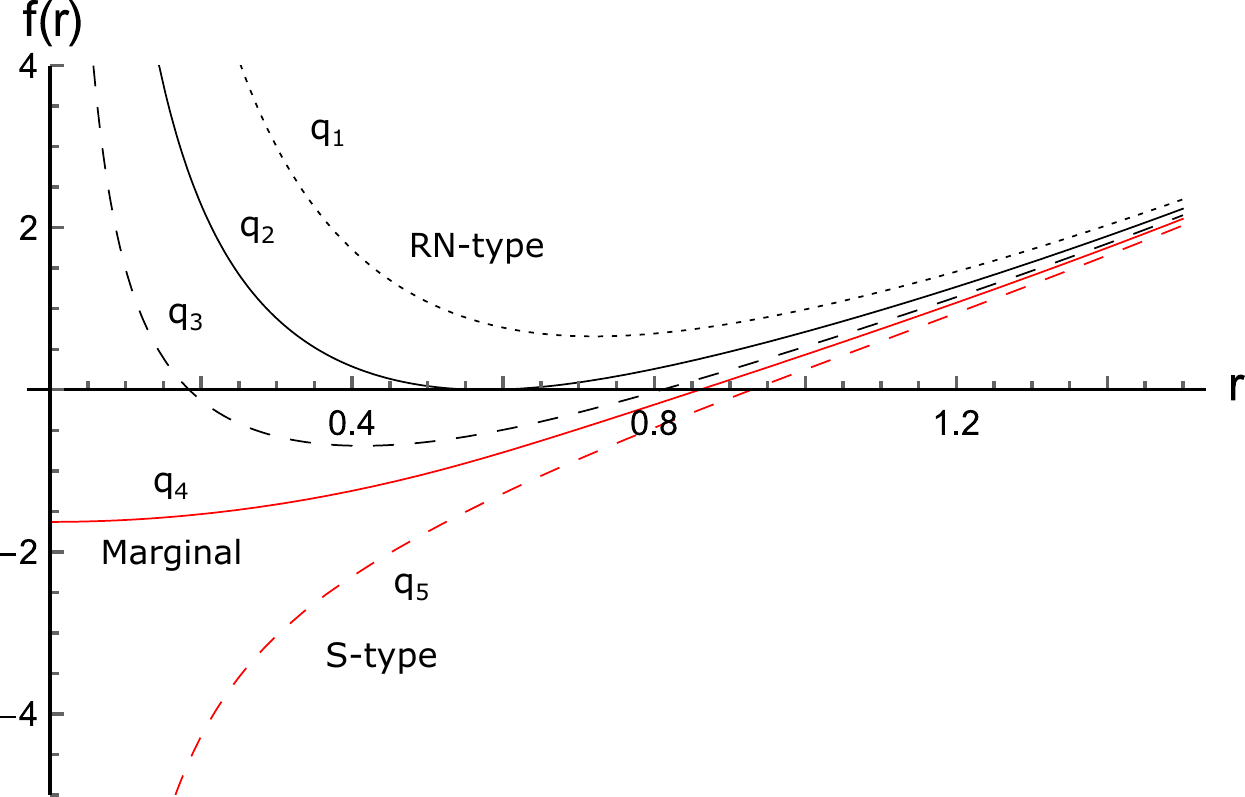}  
\caption{Behaviors of the different categories of black hole solutions for $n = 4$ with $M = 1$, $b = 2$ and $\Lambda = 0$. The black curves represent the three possible cases of RN-type solutions. The upper curve represents a solution with a naked singularity ($M < M_{ext}$). The following curves represent the extremal case ($M = M_{ext}$) and a solution with two horizons ($M~>~M_{ext}$), respectively. The upper red curve shows the marginal case, which is the one that does not present singularity in the origin. The other red curve represents a solution with a single horizon. The values of the $q_i$ that allow to obtain each curve
are displayed in table~\ref{charges-M}.}
\label{types-sol}
\end{figure}

Note that since $h(n)$ can have the indicated range of values, all these categories of solutions can be found for every $n$. In Fig.~\ref{types-sol} the first five solutions mentioned above are shown.
For this we consider the case $n = 4$ and choose $M = 1$, $b = 2$ and $\Lambda = 0$, where the values of electric charges are given according to the correspondence indicated in table~\ref{charges-M}.

\begin{table}[h!]
\centering
\resizebox{5cm}{!}{
\begin{tabular}{|l|l|} 
\hline
Electric charge $q$ &   $\times M$            \\ 
\hline
$q_1$             & 1.2                 \\ 
\hline
$q_2$ (extremal)            & $\approx 1.0000029$  \\ 
\hline
$q_3$             & 0.8                 \\ 
\hline
$q_4$             & $\approx 0.6578$      \\ 
\hline
$q_5$             & 0.5                 \\
\hline
\end{tabular}}
\caption{Values of the electric charge considered in Fig.~\ref{types-sol}.}
\label{charges-M}
\end{table}

\begin{figure}[h!]
\centering
\includegraphics[scale=0.7]{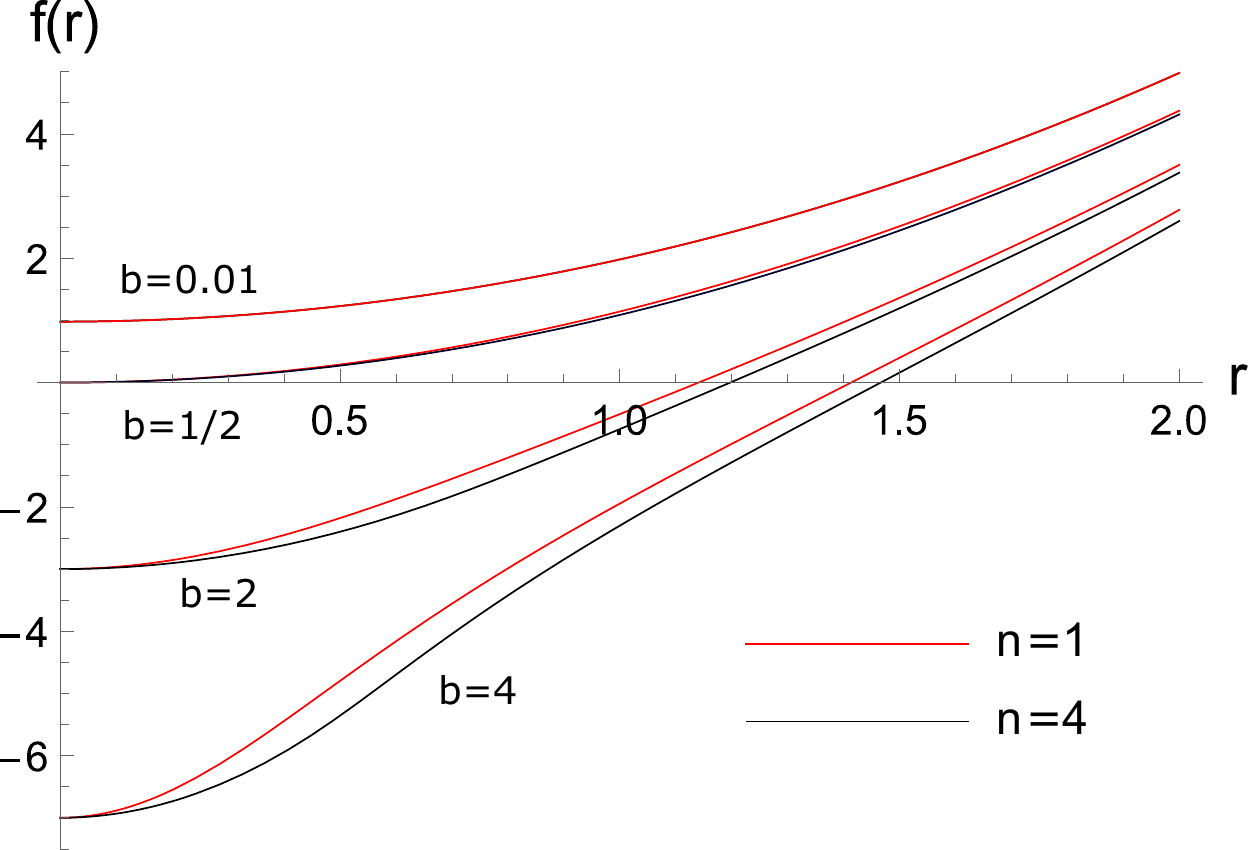}  
\caption{Comparison between the metric functions for the marginal case when $n=1$ (Born-Infeld) and $n=4$ for different values of $b$. The first two upper curves (which are closely overlapping) represent solutions with naked singularity. Meanwhile, solutions below the horizontal axis represent S-type solutions, which only have a single horizon.
We use $q=1$, $l=1$ and the values of $M = \alpha$ are obtained from Eq.~(\ref{alpha}).}
\label{marginal-1}
\end{figure}

\begin{figure}[h!]
\centering
\includegraphics[scale=0.7]{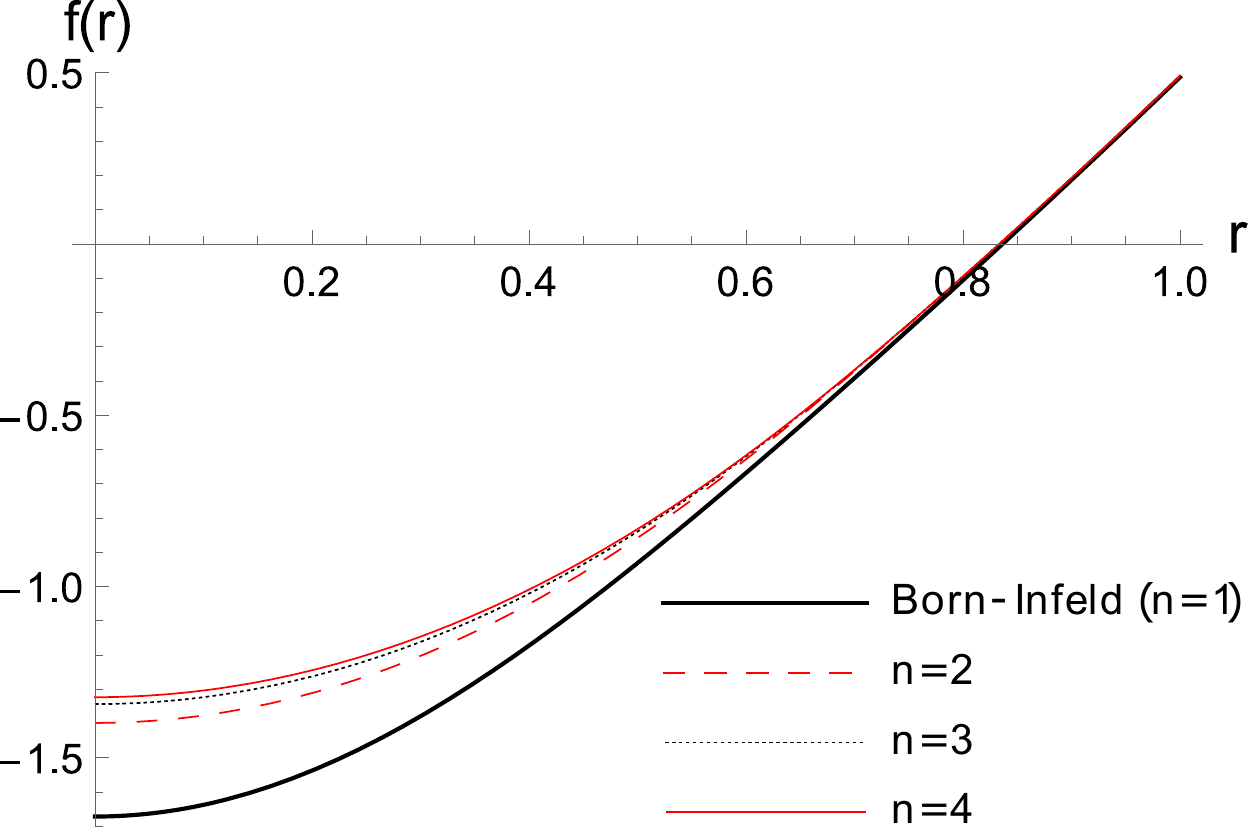}  
\caption{Again the marginal case is illustrated. Here we compare the metric functions of the Schwarzschild-type for different values of $n$, where the values of $q$ and the marginal mass remain fixed. We use $q=0.7$ and $M = \alpha =1$ and accordingly the values of the parameter $b$ vary for each $n$ as shown in table~\ref{parameter-b}.}
\label{marginal-2}
\end{figure}

In Ref.~\cite{Gunasekaran:2012dq} the parameter $\alpha$ is called marginal mass, hence the case where $M = \alpha$ is called the marginal case. In Fig.~\ref{marginal-1}  we illustrate solutions that are inside the marginal case, for different values of the parameter $b$, comparing the cases when $n=1$ and $n=4$. The solutions that intersect the vertical axis above the horizontal axis correspond to solutions with naked singularity, that is, solutions in which $bq<1/2$. Solutions such that $bq = 1/2$ are those in which the metric function at $r = 0$ is zero.
The solutions under the horizontal axis are those that behave as Schwarzschild black holes, i.e., the solutions have a single horizon. Note that if we fix the value of $bq$, then the solutions for any $n$ have the same behavior when $r \rightarrow 0$. 
Meanwhile, in Fig.~\ref{marginal-2} some metric functions are represented for different values of $n$, illustrating the marginal case for $bq > 0$, that is, solutions that have a single horizon. In this figure, in each case
the values of the mass $M = \alpha$ and $l$ are left constant, and the parameter $b$ varies according to the correspondence given in table~\ref{parameter-b}.

\begin{table}[h!]
\centering
\resizebox{2.9cm}{!}{
\begin{tabular}{|l|l|} 
\hline
$n$ &   parameter $b$            \\ 
\hline
$1$             & $1.9082    $             \\ 
\hline
$2$           & $1.7129$  \\ 
\hline
$3$             & $1.6737  $               \\ 
\hline
$4$             & $1.6593 $      \\ 
\hline
\end{tabular}}
\caption{Values of the parameter \\ $b$ used  in the graph of Fig.~\ref{marginal-2}.}
\label{parameter-b} 
\end{table}

\begin{figure}[h!]
\centering
\includegraphics[scale=0.5]{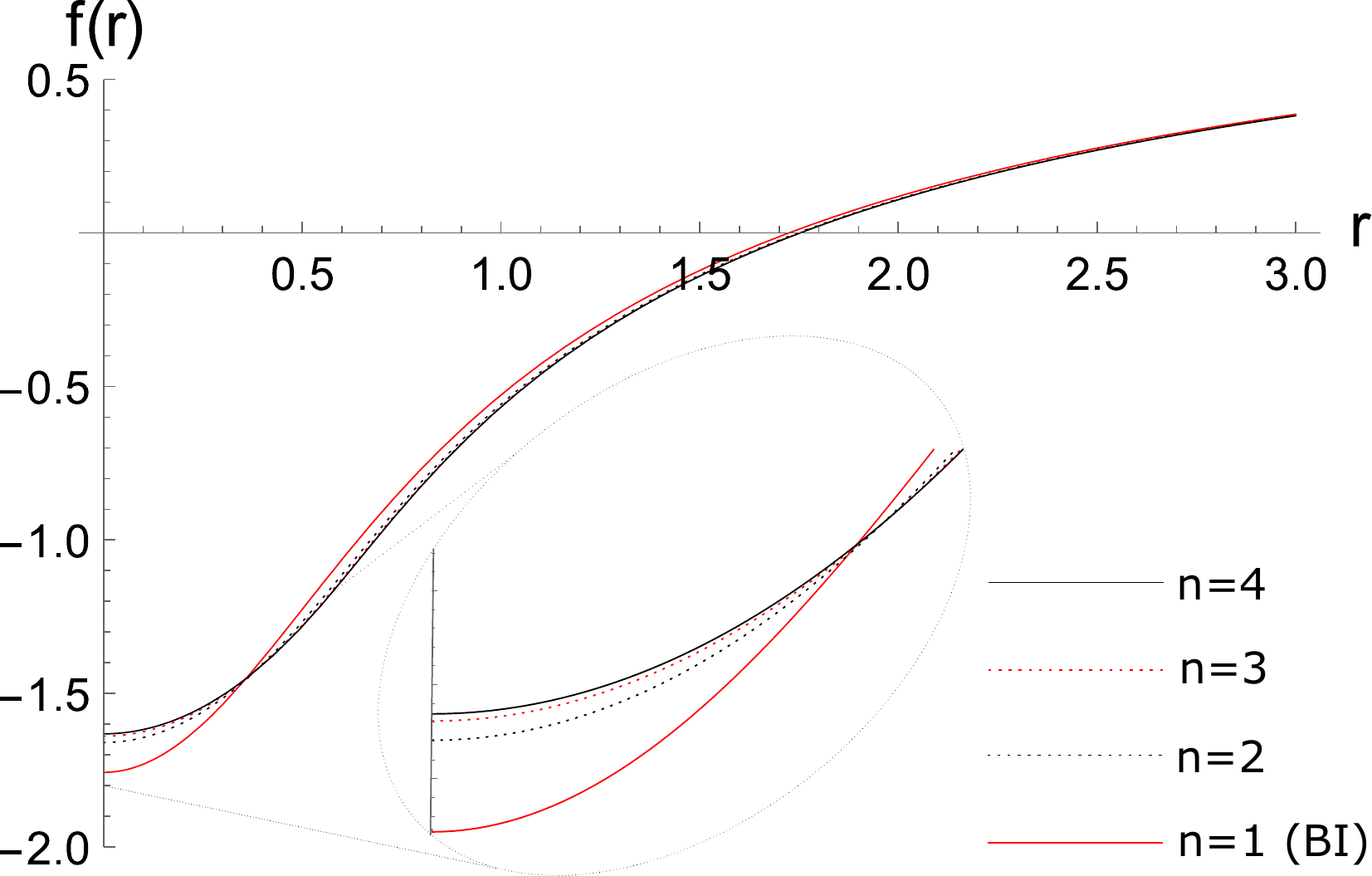}  
\caption{Marginal cases of the Schwarzschild-type for $n=1,2,3$ and $4$. 
Here we consider fixed $\Lambda=0$ and $b=2$, thus $q$ varies according to what is displayed 
in table~\ref{marginal-charge-BIt}. With these quantities, the values of $M = \alpha$ for each $n$ are obtained from Eq.~(\ref{alpha}). The oval amplifies the region where the curves intersect before reaching the vertical axis.}
\label{BIt-Marginal-Charge}
\end{figure}

\begin{table}[h]
\centering
\resizebox{4cm}{!}{
\begin{tabular}{|l|l|} 
\hline
\multicolumn{2}{|l|}{Carga eléctrica marginal}  \\ 
\hline
n & carga eléctrica q                           \\ 
\hline
1 & 0.68912714805                               \\ 
\hline
2 & 0.6647656184347                             \\ 
\hline
3 & 0.659652555844                              \\ 
\hline
4 & 0.65776354642                               \\
\hline
\end{tabular}}
\caption{Values of the electric charge that produce \\ marginal cases,  considering $\Lambda=0$, $M=1$ y $b=2$.}
\label{marginal-charge-BIt}
\end{table}
Finally, in table~\ref{marginal-charge-BIt}, we include the values of the electric charge that produce a marginal black hole for different values of $n$. This allows us to compare the behavior of the marginal case for black holes with the same mass $M=1$ and $\Lambda=0$, as shown in Fig.~\ref{BIt-Marginal-Charge}.

\vspace{.5cm}


\section{Smarr formula and first law of thermodynamics}

In this section, we will derive thermodynamic quantities, first law and the Smarr formula for the Born-Infeld-type black hole. We will be studying thermodynamics in the extended phase space where the thermodynamic pressure is given by 
\begin{equation}
P = -\frac{\Lambda}{8\pi} =\frac{3}{8 \pi l^2}
\label{pressure} \, .
\end{equation}

The Hawking temperature on the event horizon $r_+$ can be obtained by using the surface gravity $\kappa$ as
\begin{eqnarray}
T = \frac{\kappa}{2 \pi} = \frac{1}{4 \pi}  \frac{df(r_+)}{dr} &=& \frac{1}{4 \pi  r_+}
+ \frac{3 r_+}{4 \pi  l^2}
+\frac{b^2 r_+}{2 \pi} \frac{ \Gamma\left(1-\frac{1}{2 n}\right) \Gamma \left(\frac{1}{n}\right)}{\Gamma \left(\frac{1}{2 n}\right)}
\nonumber
\\&&  
   -\frac{b q}{2 \pi  r_+} \, _2F_1\left(-\frac{1}{2 n},\frac{1}{2 n};1-\frac{1}{2n};-\frac{b^{2 n} r_+^{4 n}}{q^{2 n}}\right)
\label{T-Hawking} \, .
\end{eqnarray}

The electric potential $\Phi$ on the event horizon can be calculated by using the expression for the electric field given in Eq.~(\ref{E-gral})  
\begin{equation}
\Phi(r_+) = \int_{r_+}^{\infty} E(r) dr =  \frac{\sqrt{b q} \, \Gamma\left(1 + \frac{1}{4 n}\right) \Gamma
   \left(\frac{1}{4 n}\right)}{\Gamma \left(\frac{1}{2 n}\right)}
   -b r_+
   \, _2F_1\left(\frac{1}{4 n},\frac{1}{2 n};1+\frac{1}{4 n};-\frac{b^{2 n} r_+^{4 n}}{q^{2 n}}\right)
\label{potential-E} \,  
\end{equation}

The black hole entropy is related to the event horizon as
\begin{equation}
S = \pi r_+^2
\label{entropy} \,  
\end{equation}

On the horizon $r_+$ the metric function satisfies $f(r_+) = 0$. From this one obtains the mass as a function of entropy $S$, pressure $P$, electric charge $q$ and nonlinearity parameter $b$
\begin{eqnarray}
M(S,P,q,b)  &=& \frac{b^2 S^{3/2} \Gamma \left(1-\frac{1}{2 n}\right) 
\Gamma   \left(\frac{1}{n}\right)}{3 \pi ^{3/2} \Gamma \left(\frac{1}{2
   n}\right)}+\frac{2 \sqrt{b q^3} \, \Gamma \left(1+\frac{1}{4 n}\right)
   \Gamma \left(\frac{1}{4 n}\right)}{3 \Gamma \left(\frac{1}{2
   n}\right)}+\frac{4 P S^{3/2}}{3 \sqrt{\pi }} \nonumber
\\&& +
\frac{1}{2} \sqrt{ \frac{S}{\pi }}
-\frac{b q }{3 } \sqrt{ \frac{S}{\pi }}\, _2F_1\left(-\frac{1}{2 n},\frac{1}{2
   n};1-\frac{1}{2 n};-\left(\frac{b S}{\pi  q}\right)^{2n}\right)\nonumber\\&&
   -\frac{2 b q }{3} \sqrt{ \frac{S}{\pi }}
   \, _2F_1\left(\frac{1}{4 n},\frac{1}{2
   n};1+\frac{1}{4 n};-\left(\frac{b S}{\pi  q}\right)^{2n} 
\right)
   \label{mass-SP} \, 
\end{eqnarray}

The black hole entropy is conjugate to the temperature, then 
\begin{eqnarray}
T =  \left ( \frac{ \partial M} { \partial S } \right) _{q,P,b} &=& \frac{1 }{4 \sqrt{\pi} \sqrt{S} } +
\frac{2 P \sqrt{S}}{\sqrt{\pi }}   +  \frac{b^2}{2} \sqrt{\frac{S }{\pi^3}}  \frac{  \Gamma \left(1-\frac{1}{2 n}\right) \Gamma \left(\frac{1}{n}\right)}{ \Gamma \left(\frac{1}{2n}\right)} \nonumber\\&&  -   \frac{ b q }{2 \sqrt{\pi }\sqrt{S}} \, _2F_1\left(-\frac{1}{2 n},\frac{1}{2n};1-\frac{1}{2 n};-\left(\frac{b S}{\pi  q}\right)^{2n}\right)
\label{T-M}
\end{eqnarray}

One can easily show, using Eqs.~(\ref{pressure}) and~(\ref{entropy}), that the two expressions~(\ref{T-Hawking}) and~(\ref{T-M}) for the black hole temperature are equivalent.

The electric potential on the event horizon is conjugate to the electric charge $q$, so it can also be calculated as
\begin{eqnarray}
\Phi(r_+) = \left ( \frac{ \partial M} { \partial q } \right) _{S,P,b} &=& 
\frac{\sqrt{b q} \, \Gamma \left(1+\frac{1}{4 n}\right) \Gamma
   \left(\frac{1}{4 n}\right)}{\Gamma \left(\frac{1}{2 n}\right)}
   \nonumber\\&&
   -\sqrt{\frac{S }{\pi}} b  \, _2F_1\left(\frac{1}{4 n},\frac{1}{2 n};1+\frac{1}{4 n};-\left(\frac{b S}{\pi  q}\right)^{2n}\right)
\label{potential-M} \,  .
\end{eqnarray}

Using Eq.~(\ref{entropy}), it is also easy to show that the result obtained for the electric potential with Eq.~(\ref{potential-E}) is equivalent to the one arrived at using Eq.~(\ref{potential-M}).

The differentiation of the mass $M(S,P,q,b)$, leads to the first law
\begin{equation}
dM = \left ( \frac{ \partial M} { \partial S } \right) _{q,P,b}  dS + \left ( \frac{ \partial M} { \partial P } \right) _{S,q,b} dP + \left ( \frac{ \partial M} { \partial q } \right) _{S,P, b}  dq
+ \left(\frac{\partial M}{\partial b}\right)_{S,q,P} db
\,\,\label{1-law-gral} \,  ,
\end{equation}
where the black hole volume $V$ is the conjugate quantity to pressure $P$
\begin{equation}
V = \left (\frac{ \partial M}{ \partial P } \right) _{S,q,b} 
\label{volume} \,  ,
\end{equation}
Following Refs.~\cite{Yi-Huan:2010jnv,Gunasekaran:2012dq}, for the Born-Infeld black hole, we have introduced the quantity $B$ as the conjugate quantity to $b$. Thus, we can calculate the variable $B$ as
\begin{eqnarray}
B = \left(\frac{\partial M}{\partial b}\right)_{S,q,P} &=& \sqrt{\frac{q^3}{b}}  \frac{\Gamma
   \left(1+\frac{1}{4 n}\right) \Gamma \left(\frac{1}{4 n}\right)}{3
   \Gamma \left(\frac{1}{2 n}\right)}+\frac{2 b r_+^3 \Gamma
   \left(1-\frac{1}{2 n}\right) \Gamma \left(\frac{1}{n}\right)}{3
   \Gamma \left(\frac{1}{2 n}\right)}
   \nonumber\\&&
   -\frac{2}{3} q r_+ \, _2F_1\left(-\frac{1}{2 n},\frac{1}{2
   n};1-\frac{1}{2 n};-\frac{b^{2 n} r_+^{4 n}}{q^{2 n}}\right)
      \nonumber\\&&
   -\frac{1}{3} q r_+
   \, _2F_1\left(\frac{1}{4 n},\frac{1}{2 n};1+\frac{1}{4 n};-\frac{b^{2 n} r_+^{4 n}}{q^{2 n}}\right)
\label{B-evaluated} \,  ,
\end{eqnarray}
where we have replaced $S$ using Eq.~(\ref{entropy}).

Consequently, these quantities satisfy the first law of thermodynamics 
\begin{equation}
dM = TdS +VdP + \Phi dq + B db
\,\,\label{1-law-kr} \,  .
\end{equation} 

Considering the parameter $b$ as a thermodynamic quantity helps us to obtain a Smarr formula~\cite{Smarr:1972kt} consistent with the first law of thermodynamics obtained above. Following the scaling argument given in Ref.~\cite{Kastor:2009wy}, we perform the dimensional analysis, by using the following scaling relations $S \propto L^2$,  $P \propto L^{-2}$, $q \propto L^1$ and $b \propto L^{-1}$, where $L = [\mbox{lenght}]$. Hence, as the mass is a homogeneous function of degree 1, the Euler theorem implies
\begin{equation}
M = (2)\left ( \frac{ \partial M} { \partial S } \right)   S + (-2) \left ( \frac{ \partial M} { \partial P } \right)  P + (1)\left ( \frac{ \partial M} { \partial q } \right)   q
+ (-1) \left(\frac{\partial M}{\partial b}\right) b
\,\,\label{smarr-scaling} \,  ,
\end{equation}
Thus, we can write the Smarr formula as  
\begin{equation}
M = 2 TS - 2VP + \Phi q - B b 
\,\,\label{smarr-formula} \,  .
\end{equation}

Note that using the Komar integral~\cite{Komar:1958wp}, we can find the same expression for Smarr formula. In this case, following, for example, the procedure given in Ref.~\cite{Balart:2017dzt}, we arrive at the following expression
\begin{equation}
M = \frac{\kappa A}{4 \pi} - 2 V P + \Phi q - \int_v dv  \, \omega
\,\,\label{smarr-formula-komar} \,  ,
\end{equation}
where $A$ is the surface area of the black hole, $\kappa$ is the surface gravity, $\omega = - 1/2(\mbox{trace } T)$ is the work density, $T$ represents the energy-momentum tensor of the considered electrodynamics model and the integral is evaluated outside surface of the black hole. Here it can be computed the following result
\begin{equation}
\int_v dv  \, \omega = Bb
\,\,\label{} \,  .
\end{equation}

Using a similar analysis as in Ref.~\cite{Gunasekaran:2012dq}, and since here too the parameter $b$ has units of electric field for all $n$, then $B$ can also be interpreted as polarization per unit volume, or as the work due to the nonlinearity of the electrodynamic model.


\section{Conclusions}
In this study, we have presented a family of nonlinear electrodynamics models that depend on a parameter $b$, which in all cases represents the maximum value reached by the electric field, and on a dimensionless parameter $n$ (positive integers),
that determines the different models of electrodynamics.  Each member of this family of solutions includes the invariants $\mathcal{F} = F_{ \mu \nu} F^{ \mu \nu}$ and $\mathcal{G} = F_{ \mu \nu} \tilde{F}^{ \mu \nu}$. Similar to the Born-Infeld electrodynamics, when $b \rightarrow \infty$ each model becomes Maxwell theory and the solution with $n = 1$ corresponds to the Born-Infeld model. The value of the self-energy for charged particles is finite in all cases. We have analyzed the causality and unitarity for our family of electrodynamic models considering $\mathcal{F} > 0$ and $\mathcal{F} < 0$ separately with $\mathcal{G} = 0$. Only in the cases where $n$ is even and $\mathcal{F} > 0$ the conditions are not fully satisfied.

When considering the phenomenon of birefringence neglecting the contributions of the higher orders in Eqs.~(\ref{n approx})   and~(\ref{n approx-E}), we basically obtain that all the models quantitatively present the same birefringence. That is, except for $n = 1$, in all cases the Cotton-Mouton effect is exhibited, that is, an external magnetic field induces birefringence, which is represented in the following way
\begin{equation}
\Delta n_{CM} =  n_{\parallel} -  n_{\perp}  \approx  \frac{\overline{\mathbf{B}}^2}{2 b^2}
\,\,\label{} \,  .
\end{equation}
In a similar way, we obtain that for all cases, excluding $n = 1$, the Kerr effect is present, that is, in the presence of an external electric field birefringence is induced, which is characterized as
\begin{equation}
\Delta n_{K} =  n_{\parallel} -  n_{\perp}  \approx  -\frac{\overline{\mathbf{E}}^2}{2 b^2}
\,\,\label{} \,  .
\end{equation}
According to Ref.~\cite{Battesti:2012hf} for the Cotton-Moutton and Kerr birefringences, $\Delta n_{CM} = k_{CM} \mathbf{B}^2$ and $\Delta n_ {K} = k_{K} \mathbf{E}^2$, respectively, where $k_{CM} \approx 4.03 \times 10^{-24} T^{-2}$ and $k_{K} \approx -4.4 \times 10^{-41} m^2 V^{-2}$. 

We have found a family of solutions for charged black holes coupling General Relativity with the electrodynamic models that we have presented. As in the case of a Born-Infeld black hole, we have also classified, according to their behaviour, the different solutions obtained for each $n$.
All cases exhibit the same categories, which are defined according the horizons that each solution presents.
In the transition from a Schwarzschild-type black hole to a Reissner-Nordström-type black hole, a solution (marginal case) is found whose metric function is regular at the origin, as illustrated in Figs.~\ref{marginal-1}, \ref{marginal-2} and~\ref{BIt-Marginal-Charge}. All the cases illustrated here have a horizon, except those corresponding to $b = 0.01$ in Fig.~\ref{marginal-1}.

Even though it is possible to estimate its values, the idea of considering the cosmological constant~\cite{Teitelboim:1985dp} or the nonlinearity parameter $b$~\cite{Yi-Huan:2010jnv} as dynamic variables and from this as thermodynamic quantities has been a subject of great interest~\cite {Kastor:2009wy, Gunasekaran:2012dq, Kubiznak:2016qmn}.
In this approach, the cosmological constant contributes to the first law as a thermodynamic pressure in the term $V dP$, and the nonlinearity parameter contributes as a term $B db$ that allows us to solve the problem of finding a Smarr formula consistent with the first law. The first of these quantities has been considered as a conserved charge in the form $C =- \sqrt{|\Lambda|} /4 \pi G$ (where the minus sign corresponds to the AdS case)~\cite{Chernyavsky:2017xwm} with the chemical potential $\Theta$ as its thermodynamic conjugate, contributing to the first law with the term $\Theta dC$. Perhaps also the parameter $b$ deserves equal attention from this point of view.


\section*{Acknowledgments}

L.B. was supported by DIUFRO through the project DI22-0026.






\begin{thebibliography}{00}


\bibitem{born}
M.~Born and L.~Infeld,
``Foundations of the new field theory,''
  Proc.\ Roy.\ Soc.\ Lond.\ A {\bf 144}, no. 852, 425 (1934).

\bibitem{Gibbons:2001gy}
G.~W.~Gibbons,
``Aspects of Born-Infeld theory and string / M theory,''
AIP Conf. Proc. \textbf{589}, no.1, 324-350 (2001).

\bibitem{Fradkin:1985qd}
E.~S.~Fradkin and A.~A.~Tseytlin,
``Nonlinear Electrodynamics from Quantized Strings,''
Phys. Lett. B \textbf{163}, 123-130 (1985).


\bibitem{Davila:2013wba}
J.~M.~Davila, C.~Schubert and M.~A.~Trejo,
``Photonic processes in Born-Infeld theory,''
Int. J. Mod. Phys. A \textbf{29}, 1450174 (2014).

\bibitem{Ellis:2017edi}
J.~Ellis, N.~E.~Mavromatos and T.~You,
``Light-by-Light Scattering Constraint on Born-Infeld Theory,''
Phys. Rev. Lett. \textbf{118}, no.26, 261802 (2017).

\bibitem{NiauAkmansoy:2017kbw}
P.~Niau Akmansoy and L.~G.~Medeiros,
``Constraining Born\textendash{}Infeld-like nonlinear electrodynamics using hydrogen\textquoteright{}s ionization energy,''
Eur. Phys. J. C \textbf{78}, no.2, 143 (2018).

\bibitem{NiauAkmansoy:2018ilv}
P.~Niau Akmansoy and L.~G.~Medeiros,
``Constraining nonlinear corrections to Maxwell electrodynamics using $\gamma\gamma$ scattering,''
Phys. Rev. D \textbf{99}, no.11, 115005 (2019).

\bibitem{Neves:2021tbt}
M.~J.~Neves, J.~B.~de Oliveira, L.~P.~R.~Ospedal and J.~A.~Helay\"el-Neto,
``Dispersion relations in nonlinear electrodynamics and the kinematics of the Compton effect in a magnetic background,''
Phys. Rev. D \textbf{104}, no.1, 015006 (2021).

\bibitem{Neves:2021jdy}
M.~J.~Neves, L.~P.~R.~Ospedal, J.~A.~Helay\"el-Neto and P.~Gaete,
``Considerations on anomalous photon and Z-boson self-couplings from the Born\textendash{}Infeld weak hypercharge action,''
Eur. Phys. J. C \textbf{82}, no.4, 327 (2022).

\bibitem{DeFabritiis:2021qib}
P.~De Fabritiis, P.~C.~Malta and J.~A.~Helay\"el-Neto,
``Phenomenology of a Born-Infeld extension of the $U(1)_Y$ sector at lepton colliders,''
Phys. Rev. D \textbf{105}, no.1, 016007 (2022).

\bibitem{Heisenberg:1936nmg}
W.~Heisenberg and H.~Euler,
``Consequences of Dirac's theory of positrons,''
Z. Phys. \textbf{98}, no.11-12, 714-732 (1936).

\bibitem{Kruglov:2007bh}
S.~I.~Kruglov,
``Vacuum birefringence from the effective Lagrangian of the electromagnetic field,''
Phys. Rev. D \textbf{75}, 117301 (2007).

\bibitem{Kim:2021grj}
J.~Y.~Kim,
``Deflection of light by a Coulomb charge in Born\textendash{}Infeld electrodynamics,''
Eur. Phys. J. C \textbf{81}, no.6, 508 (2021).

\bibitem{Shabad:2011hf}
A.~E.~Shabad and V.~V.~Usov,
``Effective Lagrangian in nonlinear electrodynamics and its properties of causality and unitarity,''
Phys. Rev. D \textbf{83}, 105006 (2011).


\bibitem{Soleng:1995kn} 
  H.~H.~Soleng,
``Charged black points in general relativity coupled to the logarithmic U(1) gauge theory,''
  Phys.\ Rev.\ D {\bf 52}, 6178 (1995).


\bibitem{Kruglov:2014hpa}
S.~I.~Kruglov,
``A model of nonlinear electrodynamics,''
Annals Phys. \textbf{353}, 299-306 (2014).

\bibitem{Kruglov:2014iwa}
S.~I.~Kruglov,
``Nonlinear arcsin-electrodynamics,''
Annalen Phys. \textbf{527}, 397-401 (2015).


\bibitem{Kruglov:2015fcd}
S.~I.~Kruglov,
``Modified nonlinear model of arcsin-electrodynamics,''
Commun. Theor. Phys. \textbf{66}, no.1, 59-65 (2016).


\bibitem{Kruglov:2016cdm}
S.~I.~Kruglov,
``Acceleration of Universe by Nonlinear Electromagnetic Fields,''
Int. J. Mod. Phys. D \textbf{25}, no.11, 1640002 (2016).

\bibitem{Kruglov:2017fuj}
S.~I.~Kruglov,
``Magnetized black holes and nonlinear electrodynamics,''
Int. J. Mod. Phys. A \textbf{32}, no.23n24, 1750147 (2017).

\bibitem{Kruglov:2017xmb}
S.~I.~Kruglov,
``Nonlinear Electrodynamics and Magnetic Black Holes,''
Annalen Phys. \textbf{529}, no.8, 1700073 (2017).


\bibitem{Gaete:2013dta}
P.~Gaete and J.~Helay\"el-Neto,
``Finite Field-Energy and Interparticle Potential in Logarithmic Electrodynamics,''
Eur. Phys. J. C \textbf{74}, no.3, 2816 (2014).


\bibitem{Gaete:2017cpc}
P.~Gaete and J.~A.~Helay\"el-Neto,
``A note on nonlinear electrodynamics,''
EPL \textbf{119}, no.5, 51001 (2017).

\bibitem{Gaete:2018nwq}
P.~Gaete, J.~A.~Helay\"el-Neto and L.~P.~R.~Ospedal,
``Coulomb's law modification driven by a logarithmic electrodynamics,''
EPL \textbf{125}, no.5, 51001 (2019).

\bibitem{Hendi:2012zz}
S.~H.~Hendi,
``Asymptotic charged BTZ black hole solutions,''
JHEP \textbf{03}, 065 (2012).

\bibitem{Mazharimousavi:2019sgz}
S.~H.~Mazharimousavi and M.~Halilsoy,
``Electric Black Holes in a Model of Nonlinear Electrodynamics,''
Annalen Phys. \textbf{531}, no.12, 1900236 (2019).

\bibitem{Mazharimousavi:2021uki}
S.~H.~Mazharimousavi and M.~Halilsoy,
``Electric and magnetic black holes in a new nonlinear electrodynamics model,''
Annals Phys. \textbf{433}, 168579 (2021).


\bibitem{Gullu:2020ant}
I.~Gullu and S.~H.~Mazharimousavi,
``Double-logarithmic nonlinear electrodynamics,''
Phys. Scripta \textbf{96}, no.4, 045217 (2021).



\bibitem{Gaete:2014nda}
P.~Gaete and J.~Helay\"el-Neto,
``Remarks on nonlinear Electrodynamics,''
Eur. Phys. J. C \textbf{74}, no.11, 3182 (2014).


\bibitem{Kruglov:2016uzf}
S.~I.~Kruglov,
``Notes on Born\textendash{}Infeld-type electrodynamics,''
Mod. Phys. Lett. A \textbf{32}, no.36, 1750201 (2017).


\bibitem{Bandos:2020jsw}
I.~Bandos, K.~Lechner, D.~Sorokin and P.~K.~Townsend,
``A non-linear duality-invariant conformal extension of Maxwell's equations,''
Phys. Rev. D \textbf{102}, 121703 (2020).

\bibitem{Kruglov:2021bhs}
S.~I.~Kruglov,
``On generalized ModMax model of nonlinear electrodynamics,''
Phys. Lett. B \textbf{822}, 136633 (2021).






\bibitem{Hoffmann:1935ty} 
  B.~Hoffmann,
``Gravitational and electromagnetic mass in the Born-Infeld electrodynamics,''
  Phys.\ Rev.\  {\bf 47}, no. 11, 877 (1935).

\bibitem{Hoffmann:1937noa}
B.~Hoffmann and L.~Infeld,
``On the choice of the action function in the new field theory,''
Phys. Rev. \textbf{51}, no.9, 765-773 (1937).


\bibitem{AyonBeato:1998ub} 
  E.~Ayon-Beato and A.~Garcia,
  {\it Regular black hole in general relativity coupled to nonlinear electrodynamics},
  Phys.\ Rev.\ Lett.\  {\bf 80}, 5056 (1998).

\bibitem{AyonBeato:1999rg} 
  E.~Ayon-Beato and A.~Garcia,
  {\it New regular black hole solution from nonlinear electrodynamics},
  Phys.\ Lett.\ B {\bf 464}, 25 (1999).

\bibitem{Bronnikov:2000vy} 
  K.~A.~Bronnikov,
  {\it Regular magnetic black holes and monopoles from nonlinear electrodynamics},
  Phys.\ Rev.\ D {\bf 63}, 044005 (2001).

\bibitem{Dymnikova:2004zc} 
  I.~Dymnikova,
  {\it Regular electrically charged structures in nonlinear electrodynamics coupled to general relativity},
  Class.\ Quant.\ Grav.\  {\bf 21}, 4417 (2004).


\bibitem{Balart:2014jia} 
  L.~Balart and E.~C.~Vagenas,
 {\it Regular black hole metrics and the weak energy condition},
  Phys.\ Lett.\ B {\bf 730}, 14 (2014).


\bibitem{Balart:2014cga} 
  L.~Balart and E.~C.~Vagenas,
   {\it Regular black holes with a nonlinear electrodynamics source},
  Phys.\ Rev.\ D {\bf 90}, no. 12, 124045 (2014).

\bibitem{Fernando:2003tz}
S.~Fernando and D.~Krug,
``Charged black hole solutions in Einstein-Born-Infeld gravity with a cosmological constant,''
Gen. Rel. Grav. \textbf{35}, 129-137 (2003).

\bibitem{Rasheed:1997ns}
D.~A.~Rasheed,
``Nonlinear electrodynamics: Zeroth and first laws of black hole mechanics,''
[arXiv:hep-th/9702087 [hep-th]].

\bibitem{Balart:2017dzt}
L.~Balart and S.~Fernando,
``A Smarr formula for charged black holes in nonlinear electrodynamics,''
Mod. Phys. Lett. A \textbf{32}, no.39, 1750219 (2017).


\bibitem{Gunasekaran:2012dq}
S.~Gunasekaran, R.~B.~Mann and D.~Kubiznak,
``Extended phase space thermodynamics for charged and rotating black holes and Born-Infeld vacuum polarization,''
JHEP \textbf{11}, 110 (2012).

\bibitem{Kastor:2009wy}
D.~Kastor, S.~Ray and J.~Traschen,
 {\it Enthalpy and the Mechanics of AdS Black Holes},
Class. Quant. Grav. \textbf{26}, 195011 (2009).

\bibitem{Dolan:2010ha}
B.~P.~Dolan,
``The cosmological constant and the black hole equation of state,''
Class. Quant. Grav. \textbf{28}, 125020 (2011).

\bibitem{Dolan:2011xt}
B.~P.~Dolan,
``Pressure and volume in the first law of black hole thermodynamics,''
Class. Quant. Grav. \textbf{28}, 235017 (2011).


\bibitem{Yi-Huan:2010jnv}
W.~Yi-Huan,
``Energy and first law of thermodynamics for Born-Infeld-anti-de-Sitter black hole,''
Chin. Phys. B \textbf{19}, 090404 (2010).

\bibitem{Zou:2013owa}
D.~C.~Zou, S.~J.~Zhang and B.~Wang,
``Critical behavior of Born-Infeld AdS black holes in the extended phase space thermodynamics,''
Phys. Rev. D \textbf{89}, no.4, 044002 (2014).

\bibitem{Hendi:2014kha}
S.~H.~Hendi, S.~Panahiyan and B.~Eslam Panah,
``P\textendash{}V criticality and geometrical thermodynamics of black holes with Born\textendash{}Infeld type nonlinear electrodynamics,''
Int. J. Mod. Phys. D \textbf{25}, no.01, 1650010 (2015).

\bibitem{Kumar:2019zbp}
N.~Kumar, S.~Bhattacharyya and S.~Gangopadhyay,
``Phase transitions in Born-Infeld AdS black holes in D-dimensions,''
Gen. Rel. Grav. \textbf{52}, no.2, 20 (2020).

\bibitem{Zeng:2019jta}
X.~X.~Zeng and H.~Q.~Zhang,
``Thermodynamics and weak cosmic censorship conjecture in Born-Infeld-anti-de Sitter black holes,''
Chin. Phys. C \textbf{45}, no.2, 025112 (2021).

\bibitem{Gullu:2020qni}
I.~Gullu and S.~H.~Mazharimousavi,
``Black holes in double-Logarithmic nonlinear electrodynamics,''
Phys. Scripta \textbf{96}, no.9, 095213 (2021).

\bibitem{Balart:2021glm}
L.~Balart and S.~Fernando,
``Thermodynamics and heat engines of black holes with Born\textendash{}Infeld-type electrodynamics,''
Mod. Phys. Lett. A \textbf{36}, no.15, 15 (2021).

\bibitem{Kumar:2022fyq}
N.~Kumar, S.~Sen and S.~Gangopadhyay,
``Phase transition structure and breaking of universal nature of central charge criticality in a Born-Infeld AdS black hole,''
Phys. Rev. D \textbf{106}, no.2, 026005 (2022).



\bibitem{Beard:2021dka}
J.~B\'eard, J.~Agil, R.~Battesti and C.~Rizzo,
``A novel pulsed magnet for magnetic linear birefringence measurements,''
Rev. Sci. Instrum. \textbf{92}, no.10, 104710 (2021).

\bibitem{Dehghani:2021fwb}
A.~Dehghani, M.~R.~Setare and S.~Zarepour,
``Self-energy problem, vacuum polarization, and dual symmetry in Born\textendash{}Infeld-type U(1) gauge theories,''
Eur. Phys. J. Plus \textbf{137}, no.7, 859 (2022)
[erratum: Eur. Phys. J. Plus \textbf{137}, no.9, 1090 (2022)].




\bibitem{Battesti:2012hf}
R.~Battesti and C.~Rizzo,
``Magnetic and electric properties of quantum vacuum,''
Rept. Prog. Phys. \textbf{76}, no.1, 016401 (2013).

\bibitem{Boillat:1970gw}
G.~Boillat,
``Nonlinear electrodynamics - Lagrangians and equations of motion,''
J. Math. Phys. \textbf{11}, no.3, 941-951 (1970).


\bibitem{Bialynicka-Birula:1970nlh}
Z.~Bialynicka-Birula and I.~Bialynicki-Birula,
``Nonlinear effects in Quantum Electrodynamics. Photon propagation and photon splitting in an external field,''
Phys. Rev. D \textbf{2}, 2341-2345 (1970).


\bibitem{Gaete:2021ytm}
P.~Gaete and J.~A.~Helay\"el-Neto,
``Remarks on inverse electrodynamics,''
Eur. Phys. J. C \textbf{81}, no.10, 899 (2021).


\bibitem{Gaete:2015qda}
P.~Gaete,
``Some Remarks on Nonlinear Electrodynamics,''
Adv. High Energy Phys. \textbf{2016}, 2463203 (2016).

\bibitem{Sorokin:2021tge}
D.~P.~Sorokin,
``Introductory Notes on Non-linear Electrodynamics and its Applications,''
Fortsch. Phys. \textbf{70}, no.7-8, 2200092 (2022).



\bibitem{Kruglov:PLA}
S.~I.~Kruglov,
``Nonlinear electrodynamics with birefringence,
Phys. Lett. B \textbf{379}, 623-625 (2015).





\bibitem{Fernando:2006gh}
S.~Fernando,
``Thermodynamics of Born-Infeld-anti-de Sitter black holes in the grand canonical ensemble,''
Phys. Rev. D \textbf{74}, 104032 (2006).

\bibitem{Breton:2017hwe}
N.~Bret\'on, T.~Clark and S.~Fernando,
``Quasinormal modes and absorption cross-sections of Born\textendash{}Infeld\textendash{}de Sitter black holes,''
Int. J. Mod. Phys. D \textbf{26}, no.10, 1750112 (2017).

\bibitem{Smarr:1972kt}
  L.~Smarr,
 ``Mass formula for Kerr black holes,''
  Phys.\ Rev.\ Lett.\  {\bf 30}, 71 (1973).
   Erratum: [Phys.\ Rev.\ Lett.\  {\bf 30},  521 (1973)].

\bibitem{Komar:1958wp}
A.~Komar,
``Covariant conservation laws in general relativity,''
Phys. Rev. \textbf{113}, 934-936 (1959).
  
\bibitem{Teitelboim:1985dp}
C.~Teitelboim,
``The cosmological constant as a thermodynamic black hole parameter,''
Phys. Lett. B \textbf{158}, 293-297 (1985).
  
\bibitem{Kubiznak:2016qmn}
D.~Kubiznak, R.~B.~Mann and M.~Teo,
``Black hole chemistry: thermodynamics with Lambda,''
Class. Quant. Grav. \textbf{34}, no.6, 063001 (2017).

\bibitem{Chernyavsky:2017xwm}
D.~Chernyavsky and K.~Hajian,
``Cosmological constant is a conserved charge,''
Class. Quant. Grav. \textbf{35}, no.12, 125012 (2018).
  
  



\end{thebibliography}
\end{document}